\documentclass[aps,pra,reprint,showpacs,amsmath,amssymb,superscriptaddress]{revtex4-1}

\usepackage{graphicx}
\usepackage{epstopdf}
\usepackage{dcolumn}
\usepackage{bm}
\usepackage{tabularx}
\usepackage{multirow}

\begin{document}

\preprint{}

\title{Optical Spectroscopy of Tungsten Carbide for Uncertainty Analysis\\ in Electron Electric Dipole Moment Search}

\author{J. Lee}

\affiliation{%
Department of Physics, University of Michigan, Ann Arbor, Michigan 48109-1040, USA
}%

\author{J. Chen}

\affiliation{%
Department of Physics, University of Michigan, Ann Arbor, Michigan 48109-1040, USA
}%

\author{L.V.\ Skripnikov}
\author{A.N.\ Petrov}
\affiliation{Federal state budgetary institution ``Petersburg Nuclear Physics  Institute'', Gatchina, Leningrad district 188300, Russia}
\affiliation{
Quantum Mechanics Division,
Saint Petersburg State University, Saint Petersburg, Petrodvoretz 198904, Russia}
\author{A.V.\ Titov}
\author{N.S.\ Mosyagin}
\affiliation{Federal state budgetary institution ``Petersburg Nuclear Physics  Institute'', Gatchina, Leningrad district 188300, Russia}

\author{A.E. Leanhardt}

\affiliation{%
Department of Physics, University of Michigan, Ann Arbor, Michigan 48109-1040, USA
}%


\begin{abstract}
We perform laser induced fluorescence(LIF) spectroscopy on a pulsed supersonic beam of tungsten carbide(WC) molecules, which has been proposed as a candidate molecular system for a permanent Electric Dipole Moment(EDM) search of the electron in its rovibrational ground state of the $X^3\Delta_1$ state. In particular, $[20.6]\Omega=2, v'=4 \leftarrow X^3\Delta_1,v"=0$ transition at 485nm was used for the detection. The hyperfine structure and the $\Omega$-doublet of the transition are measured, which are essential for estimating the size of the potential systematic uncertainties for electron EDM measurement. For further suppression of the systematic uncertainty, an alternative electron EDM measurement scheme utilizing the g factor crossing point of the $\Omega$-doublet levels is discussed. On the other hand, flux and internal temperature of the molecular beam are characterized, which sets the limit on the statistical uncertainty of the electron EDM experiment. With the given results, the prospect of electron EDM experiment with the $X^3\Delta_1$ state of WC molecule is discussed.

\end{abstract}

\maketitle

\section{\label{sec:level1}Introduction}

Physics beyond the Standard Model have been developed for many years, along with the experimental efforts ranging from the Large Hadron Collider to laboratory-based tabletop precision measurements~\cite{Jung2009}. CP violation in the Standard Model predicts an electron to have a permanent Electric Dipole Moment(EDM) of $|d_e| \approx 10^{-38}$~e-cm, however, various extensions of the Standard Model predict electron EDMs of 10 orders of magnitude larger~\cite{Pospelov2005}. Therefore, a non-zero measurement of the electron EDM becomes a direct observation of physics beyond the Standard Model. On the other hand, a null measurement of electron EDM with a small enough uncertainty would serve as a constrain for theoretical models.

The valence electrons of a suitably chosen diatomic molecule have been known to have benefits for the electron EDM search~\cite{Kozlov1995}. The EDM of the valence electron gives rise to an energy splitting between opposite spin parity states when the molecule is polarized with an laboratory electric field, $\mathcal{E_{\textrm{lab}}}$. This energy splitting is proportional to the effective electric field experienced by the electron, $\mathcal{E_{\textrm{eff}}}$. There is a big gain of $\mathcal{E_{\textrm{eff}}}$ in heavy polar molecules compared to atoms~\cite{Kozlov1995}, which opened up a new generation of molecule based electron EDM experiments~\cite{Hudson2011, Vutha2010, Meyer2008, ShaferRay2006, Lee2009}. In particular, the ytterbium fluoride (YbF) system reported a new experimental limit on the electron EDM to be $|d_e| < 1.05 \times 10^{-27}$~e-cm~\cite{Hudson2011}.

\begin{figure}
\includegraphics[width = 3.3 in]{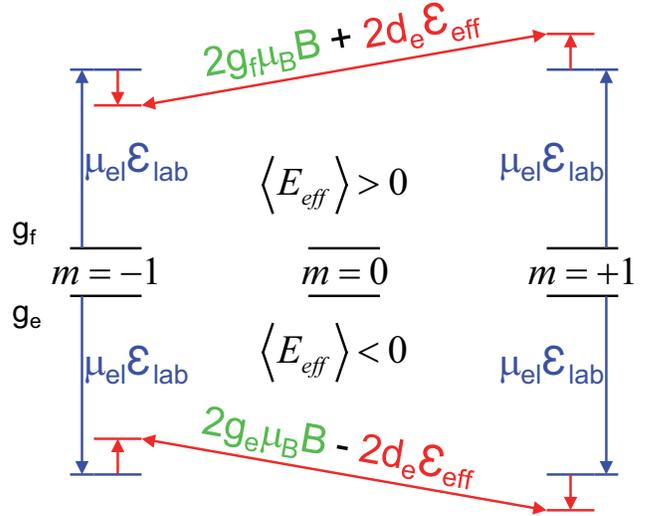}
\caption{Diagram of Zeeman shift cancelation scheme in WC electron EDM experiment. The tensor Stark shift $\mu_{el}\mathcal{E_{\textrm{lab}}}$, Zeeman shift $g\mu_BB$, and the eEDM Stark shift $d_e\mathcal{E_{\textrm{eff}}}$ are shown.}
\label{f:WCUncertainty}
\end{figure}

Recently, the $X^3\Delta_1$ ground state of tungsten carbide(WC) has been proposed as another candidate system for electron EDM measurement~\cite{Lee2009}. This system carries two main advantages for measuring the electron EDM. The first advantage comes from the large $\mathcal{E_{\textrm{eff}}}$ being applied to the valence electrons of the molecule, which is due to the heavy tungsten nucleus. The second advantage is the use of internal comagnetometer~\cite{Lee2009} illustrated in Fig.~\ref{f:WCUncertainty}, where we can use the closely spaced $\Omega$-doublet levels for Zeeman shift cancelation. When $g_e$ and $g_f$ are close enough, we effectively cancel out the Zeeman shift terms in Fig.~\ref{f:WCUncertainty} by taking the difference between the energy shifts of top and bottom measurements, and only leave the Stark shift term, which is proportional to $d_e$.

On the other hand, there are also uncertainties for two of these aspects and therefore careful analysis is required. As for the $\mathcal{E_{\textrm{eff}}}$, since there is no direct way of measuring the electric field inside the molecule, the uncertainty purely comes from the calculation itself. The calculation of $\mathcal{E_{\textrm{eff}}}$ requires information on electron wavefuction at the heavy nucleus of the molecule~\cite{Sandars1966}, therefore being strongly related to the hyperfine structure. In other words, comparing the calculated hyperfine constant with the experimental result could be a qualitative test of self consistency in theoretical calculation of the $\mathcal{E_{\textrm{eff}}}$.

The Zeeman shift cancelation scheme shown in Fig.~\ref{f:WCUncertainty} also has room for systematic uncertainty, as the small difference in g-factors between the top and bottom doublet would result in imperfect cancelation. The magnitude of this systematic uncertainty can be written as, $2\mu _{B}B\Delta g$, where $\mu _{B}$ is the Bohr magneton, B is the laboratory magnetic field, and $\Delta g$ is the difference between $g_e$ and $g_f$. Petrov~\cite{Petrov2011} has shown that $\Delta g$ is closely related to the energy splitting between top and bottom levels of the $\Omega$-doublet. Utilizing this relation, we can calculate $\Delta g$ based on $\Omega$-doublet constant measurement, and assign systematic uncertainty for electron EDM measurement.

In order to analyze these uncertainties, here we have studied the $[20.6]\Omega=2, v'=4 \leftarrow X^3\Delta_1,v"=0$ transition of WC by laser induced fluorescence(LIF) spectroscopy on a pulsed supersonic jet of WC molecules. The optical properties of WC have been studied both theoretically~\cite{Balasubramanian2000,Stevens2006} and experimentally~\cite{Sickafoose2002,Rothgeb2008,Wang2011a,Wang2011b,Wang2012}. We had $\sim100$ times higher spectral resolution than the previous resonant two-photon ionization(R2PI) spectroscopy of the same transition~\cite{Sickafoose2002}. The rotational spectrum of $[20.6]\Omega=2, v'=4 \leftarrow X^3\Delta_1,v"=0$ transition at 485nm was presented for the low lying R lines($\Delta J = +1$). The magnetic hyperfine constants and the $\Omega$-doublet constants of ground and excited state of the transition are reported and analyzed closely related to the systematic uncertainties of the electron EDM measurement scheme. The flux and internal temperature of the beam are characterized, which sets the limit on statistical sensitivity of the electron EDM experiment.

\section{Experimental Methods}

We use Smalley type pulse supersonic beam technique~\cite{Dietz1981} to generate WC beam with low rotational temperature. The requirement of low rotational temperature is not only to simplify the rotational spectrum, but also to enhance the statistics for the electron EDM experiment, which will be discussed in section V.

\begin{figure}
\includegraphics[width = 3.3 in]{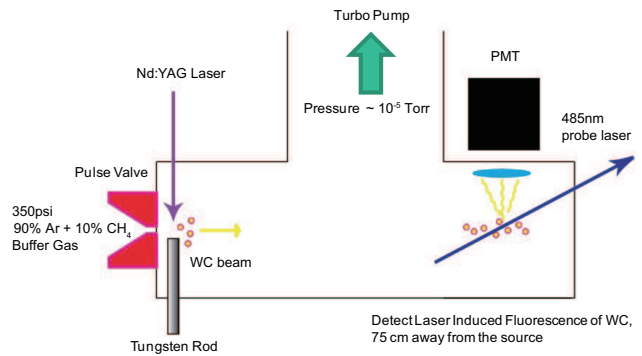}
\caption{Diagram of tungsten carbide beam apparatus.}
\label{f:Vacuum}
\end{figure}

Tungsten atoms are ablated from a rod (American Elements, $99.9\%$ purity) by the third harmonic of the Nd:YAG pulse laser (Quantel), while the solenoid gas valve (Parker, general valve series 999) entrains the atoms with 350psi of Argon buffer gas pressure. The WC molecules were generated by adding a small fraction of methane to the buffer gas, which allows for the chemical reaction $\textrm{W}+\textrm{CH}_4 \rightarrow \textrm{WC} + 2\textrm{H}_2$ to happen. The molecules get cooled down through buffer gas collisions, resulting low internal temperatures. The turbo pump with $1500$L/s of pumping speed maintained the operating pressures to be, $5\times 10^{-6}$ Torr inside the vacuum chamber. Diagram of the experimental apparatus is shown in Fig.~\ref{f:Vacuum}.

A tunable CW diode laser(Toptica DL pro) at the wavelength of 484nm-487nm range covered all the transitions presented in this paper. The probe laser was focused with an intensity of $\sim$ $100mW/cm^2$ at the intersection point where the laser beam is crossing the molecular beam perpendicularly. The laser induced fluorescing light was collected by a spherical lens into a water cooled Photo Multiplying Tube(Hammamatsu) connected to the photon counter. A $485\pm1.5$nm bandpass filter was installed in front of the PMT to only let the $[20.6]\Omega=2, v'=4 \rightarrow X^3\Delta_1,v"=0$ fluorescence to pass through. The photon counts were recorded simultaneously as the wavelength meter(High Finesse WSU series), which is a Fizeau interferometer with a frequency reference, measures the frequency of the probe laser. The uncertainty is defined in two different ways, one is in terms of absolute frequency, and the other is the relative frequency shift.

\begin{figure}
\includegraphics[width = 3.3 in]{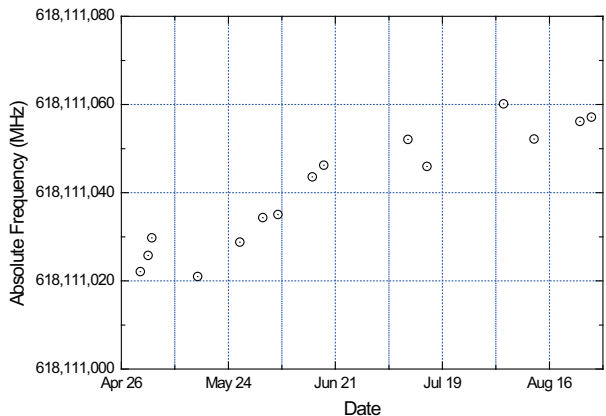}
\caption{The $^{182}$W$^{12}$C, R(1) line position of $[20.6]\Omega=2, v'=4 \leftarrow X^3\Delta_1,v"=0$ transition has been measured over 4 months period.}
\label{f:WCDrift}
\end{figure}

The uncertainty of absolute frequency measurement depends on the stability of the frequency reference which the wavelength meter synchronizes to. The current settings use a frequency stabilized HeNe laser(SIOS-02 series) with a stability of $\sim10 MHz$, which was claimed by the manufacturer. In other words, there could be an inconsistency at the same level in absolute frequency measurement. We show on Fig.~\ref{f:WCDrift}, where the absolute frequency of $^{182}$W$^{12}$C, R(1) $[20.6]\Omega=2, v'=4 \leftarrow X^3\Delta_1,v"=0$ transition was measured over a period of several months to show this effect. From the plot, we saw an $1 \sigma$ uncertainty of about $20$MHz over 4 month period. This is assigned as a systematic error in absolute frequency.

What we are more interested in is the uncertainty of relative frequency shift, as our main concern is to measure the difference in energy levels coming from various interaction Hamiltonians of WC. If we were to measure a relative frequency shift between two LIF lines, the uncertainty of each line positions explained above would become irrelevant, as it gets canceled out when we take the difference in frequency measurements. Therefore we are only left with the uncertainty that comes from the interferometer part of the wavelength meter, where we have assigned $1.6$MHz of statistical uncertainty and $6\times10^{-4}$ of fractional systematic uncertainty for our relative frequency shift measurements (see the Appendix of \cite{Lee2012a}). All of these uncertainties are listed on table~\ref{table:1}. Both statistical and systematic uncertainties are taken into account for $1\sigma$ errors of our experimental results in the following section.

\begin{table}
\caption{Experimental sources of error in frequency measurement.}\label{table:1}

\begin{ruledtabular}
\begin{tabular}{@{}lr}
Source of Error & Estimate \\
\hline
&\\
Statistical Error & 1.6MHz \\
&\\
Fractional Systematic Error &\\
in Relative Frequency& $6\times10^{-4}$\\
&\\
Systematic Error in &\\
Absolute Frequency& 20MHz\\
&\\

\end{tabular}
\end{ruledtabular}
\end{table}

\section{Experimental Results}

\subsection{Measured WC Transitions}

\begin{figure}
\includegraphics[width = 3.3 in]{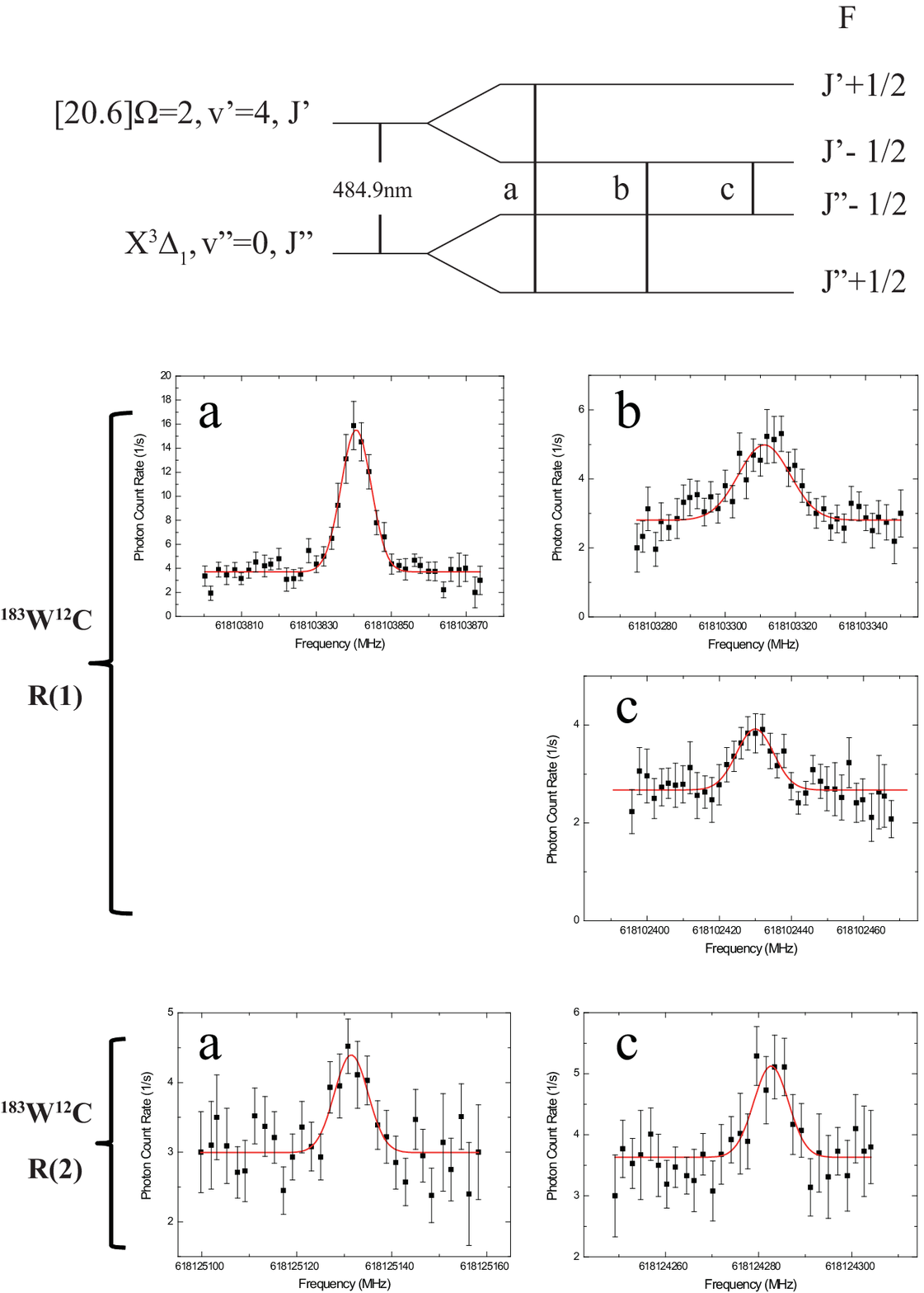}
\caption{R(1) and R(2) lines of $[20.6]\Omega=2, v'=4 \leftarrow X^3\Delta_1,v"=0$ transition was measured for $^{183}$W$^{12}$C. The photon count rates are averaged over 20 second of integration time. 1$\sigma$ error bars are shown for the photon count rates and the red solid line corresponds to the least square Gaussian fit.}
\label{f:183WCHyper}
\end{figure}

Tungsten has four major isotopes of $^{182}$W, $^{183}$W, $^{184}$W, $^{186}$W, while Carbon has only one major isotope of $^{12}$C with close to $99\%$ abundance. The $^{180}$W isotope has a natural abundance of 0.12$\%$, which was not detected with our signal to noise level. As a result, four WC isotopes are observed for each rotational transitions. The LIF signal was detected with $1\sim10$ photon counts per second, which we have averaged over 20 second of integration time per data point to get better signal to noise. The background level is dominated by the randomly scattered probe laser light going into the PMT. The strength of the signal varied on daily basis at $30\%$ level, depending on the optimization of the WC beam quality.

\begin{figure}
\includegraphics[width = 3.3 in]{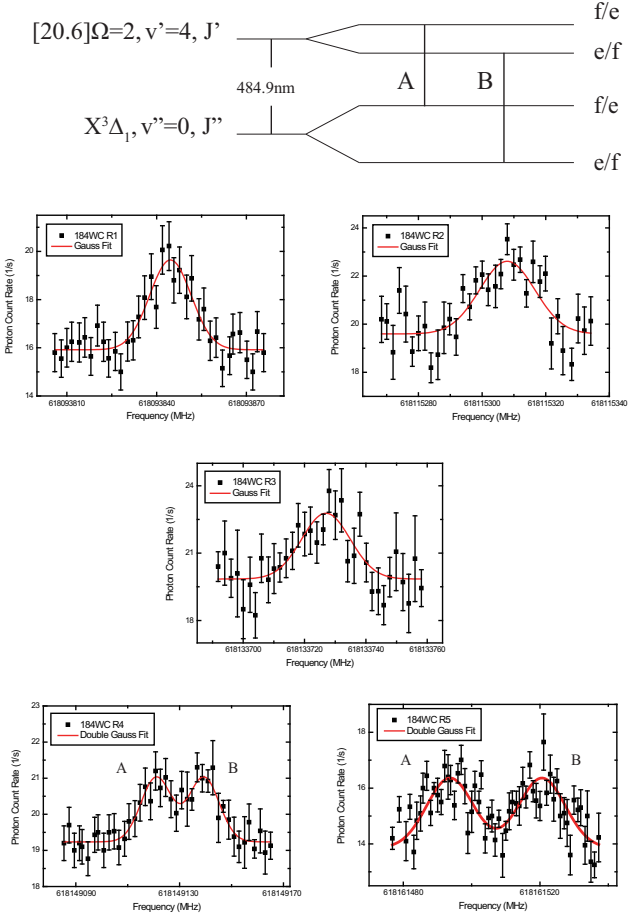}
\caption{R(1) $\sim$ R(5) lines of $[20.6]\Omega=2, v'=4 \leftarrow X^3\Delta_1,v"=0$ transition for $^{184}$W$^{12}$C are shown. The photon count rates are averaged over 20 second of integration time. 1$\sigma$ error bars are shown for the photon count rates and the red solid line corresponds to the least square Gaussian fit.}
\label{f:WCDoublet}
\end{figure}

We measured the R branch ($\Delta J = +1$) of $[20.6]\Omega=2, v'=4 \leftarrow X^3\Delta_1,v"=0$ transition at 485nm, where the results of $^{183}$W$^{12}$C in Fig.~\ref{f:183WCHyper} shows the Hyperfine structure and the results of $^{184}$W$^{12}$C in Fig.~\ref{f:WCDoublet} shows the $\Omega$-doublet structure. We observed the same $\Omega$-doublet structure in $^{182}$W$^{12}$C and $^{186}$W$^{12}$C isotopes as well, which will be shown in the Appendix section.

Each of the individual line had a Doppler Broadened linewidth of $\sim10MHz$. Least square Gaussian fit was used to extract the center frequency of the line. The fitting error for the center frequency was only in the order of sub MHz, therefore, the uncertainties from the wavelength meter discussed in the previous section were the dominating sources of error. The list of our measured line positions are shown with proper error assignments and compared with the previous results of ref.~\cite{Sickafoose2002} on Table~\ref{table:2}. The center of gravity position is shown for the $^{183}$W$^{12}$C isotope, as it has a hyperfine structure. For the $\Omega$-doublet structure observed in R(4) and R(5) transitions, the center frequency between the double peak is listed on the table. Sickafoose's data shows the fitted line positions with the residuals from the fit given in the parentheses in the units of last significant digit.

The $^{183}$W$^{12}$C isotope has a non-zero nuclear spin, which gives rise to hyperfine structure. Only the magnetic dipole interaction is present as it has a nuclear spin of $1/2$~\cite{Kopfermann1958}. From the selection rule, there are three allowed electric dipole transitions of $\Delta F = 0,\pm1$ for each R lines. We have measured 5 out of 6 hyperfine transitions in $^{183}$W$^{12}$C, R(1) and R(2) lines, which are shown on Fig.~\ref{f:183WCHyper}. The $^{183}$W$^{12}$C isotope has a relatively low natural abundance of $14\%$ compared to other isotopes, therefore, only up to R(2) transition was detected. Also, the $^{183}$W$^{12}$C, b transition of R(2) line has a relatively small Clebsch-Gordan coefficient which made it undetectable with our signal to noise. The intensity ratios of these transitions are calculated to be $[I_a : I_b : I_c = 9 : 1 : 5]$ for R(1) and $[I_a : I_b : I_c = 20 : 1 : 14]$ for R(2).

\begin{table}
\caption{Absolute frequencies of measured lines in $[20.6]\Omega=2, v'=4 \leftarrow X^3\Delta_1,v"=0$ transition are compared with ref.~\cite{Sickafoose2002}. Center of gravity position is shown for $^{183}$W$^{12}$C isotopes. As for the R(4) and R(5) transitions, we start to see an $\Omega$-doublet structure, where the center frequencies between the peaks are listed. The 1 $\sigma$ errors are shown inside the parentheses in the order of first parenthesis with the statistical uncertainty, and the second parenthesis with the systematic uncertainty in absolute frequencies. Only the 1 $\sigma$ fitting error was provided from ref.~\cite{Sickafoose2002}}\label{table:2}

\begin{ruledtabular}
\begin{tabular}{@{}lrr}

Measured Line &This Work & Ref.~\cite{Sickafoose2002}\\
               &(MHz)     &(MHz)\\
\hline
&&\\
$^{182}$W$^{12}$C, R(1)&618,110,996(1.6)(20)&618,110,980(690)\\
$^{182}$W$^{12}$C, R(2)&618,132,466(1.6)(20)&618,132,416(540)\\
$^{182}$W$^{12}$C, R(3)&618,150,892(1.6)(20)&618,151,362(-180)\\
$^{182}$W$^{12}$C, R(4)&618,166,298(1.6)(20)&618,166,802(-480)\\
$^{182}$W$^{12}$C, R(5)&618,178,709(1.6)(20)&618,178,613(-270)\\
&&\\
$^{183}$W$^{12}$C, R(1)&618,103,338(1.6)(20)&618,102,617(-30)\\
$^{183}$W$^{12}$C, R(2)&618,124,778(1.6)(20)&618,123,992(-30)\\
&&\\
$^{184}$W$^{12}$C, R(1)&618,093,845(1.6)(20)&618,093,443(540)\\
$^{184}$W$^{12}$C, R(2)&618,115,308(1.6)(20)&618,115,208(180)\\
$^{184}$W$^{12}$C, R(3)&618,133,727(1.6)(20)&618,133,345(390)\\
$^{184}$W$^{12}$C, R(4)&618,149,130(1.6)(20)&618,149,624(-570)\\
$^{184}$W$^{12}$C, R(5)&618,161,507(1.6)(20)&618,162,095(-780)\\
&&\\
$^{186}$W$^{12}$C, R(1)&618,077,509(1.6)(20)&618,076,625(660)\\
$^{186}$W$^{12}$C, R(2)&618,098,953(1.6)(20)&618,098,509(240)\\
$^{186}$W$^{12}$C, R(3)&618,117,361(1.6)(20)&618,117,186(-30)\\
$^{186}$W$^{12}$C, R(4)&618,132,760(1.6)(20)&618,132,805(-210)\\
$^{186}$W$^{12}$C, R(5)&618,145,138(1.6)(20)&618,145,397(-420)\\
&&\\
\end{tabular}
\end{ruledtabular}
\end{table}

Figure~\ref{f:WCDoublet} shows the five lowest R lines of the $^{184}$W$^{12}$C isotope for the given transition. Both the X$^3\Delta_1$ state and the $[20.6]\Omega=2$ have nearly degenerate spin states of opposite parity called the $\Omega$-doublet, where there are two allowed electric dipole transitions among them for each R transitions. Following the notations of Brown~\cite{Brown1987}, we label them $e/f \leftrightarrow e/f$. Due to the J dependence of the doublet interaction, $H_{doublet}\propto J(J+1)$, we only see the doublet structure when the interaction term becomes larger than our Doppler limited linewidth. As a result, we saw the doublet structure only at higher J line transitions, which are R(4) and R(5) transitions.

\subsection{Hyperfine Constant}

As discussed previously, the hyperfine structure of $^{183}$W$^{12}$C isotope is only cause by magnetic dipole interaction, which can be written as,

\begin{equation}\label{eq:2}
H=h\times \frac{F(F+1)-J(J+1)-I(I+1)}{2J(J+1)},
\end{equation}
where $h$ is the magnetic hyperfine constant. The hyperfine frequency splitting within the electronic state becomes,

\begin{equation}\label{eq:3}
\delta\nu=h\left ( \frac{J+\frac{1}{2}}{J(J+1)} \right ).
\end{equation}

It is straightforward to extract the magnetic hyperfine constants of ground and excited state of the transition using equation~\ref{eq:3}, with given hyperfine frequency splittings among the peaks of Fig.~\ref{f:183WCHyper}. First we had to assign each measured frequency splittings to the correct hyperfine transition (i.e. whether the splitting corresponds to $\left |a-b\right |$, $\left |b-c\right |$, or $\left |c-a\right |$ transition). We have used the least square fitting method for all combinations of assignments, and picked the assignment which gave us the smallest residuals for the fit. Using the wrong assignment made a significant increase in the residuals, which allowed us to reject it. The result of three independent frequency splittings with the correct assignments are shown on table~\ref{table:4}. Based on this assignment, we report the measured hyperfine constants to be $h_{3\Delta_1,v"=0} = -1171(4)$MHz and $h_{[20.6]\Omega=2,v'=4} = 1258(6)$MHz with $1\sigma$ fitting errors shown inside the parentheses. Detailed comparison with the calculated hyperfine constant will be shown in section IV.

\begin{table}
\caption{Relative frequency shifts of measured hyperfine splittings of $^{183}$W$^{12}$C in $[20.6]\Omega=2, v'=4 \leftarrow X^3\Delta_1,v"=0$ transition. The 1 $\sigma$ errors are shown inside the parentheses in the order of first parenthesis with the statistical uncertainty, and the second parenthesis with the systematic uncertainty in relative frequencies. The residuals from the fit are shown on the third column.}\label{table:4}

\begin{ruledtabular}
\begin{tabular}{@{}lrr}

Measured Line &Splitting & Fit Residual\\
               &(MHz)     &(MHz)\\
\hline
&&\\
$^{183}$W$^{12}$C, R(1), $\left |a-b\right |$ &528(1.6)(0.3)&-2.3\\
$^{183}$W$^{12}$C, R(1), $\left |b-c\right |$ &882(1.6)(0.5)&-1.5\\
$^{183}$W$^{12}$C, R(2), $\left |a-c\right |$ &849(1.6)(0.5)&-9.8\\
&&\\
\end{tabular}
\end{ruledtabular}
\end{table}

\subsection{$\Omega$-doublet Constant}

Both X$^3\Delta_1$ state and $[20.6]\Omega=2$ state of WC have parity eigenstates in each of its $J$ rotational levels, which are known as the $\Omega$-doublet. These doublet structures have been studied by Brown~\cite{Brown1987}, where they have proposed a labeling convention of $|e\rangle$ levels for parity $+(-1)^J$ and $|f\rangle$ levels for parity $-(-1)^J$. The coupling between the rotational and electronic motion lifts the degeneracy of the parity eigenstates, causing energy splittings for states with $\Omega \neq 1$. Following ref.~\cite{Brown1987}, we write down the Hamiltonian for both states as,

\begin{eqnarray}\label{eq:4}
&H_{^3\Delta_1}=\pm \tilde{o}_{^3\Delta_1}J(J+1),\\
&H_{[20.6]\Omega =2}=\pm \tilde{o}_{[20.6]\Omega =2}(J-1)J(J+1)(J+2),
\end{eqnarray}
where $\tilde{o}_{^3\Delta_1}$ $\tilde{o}_{[20.6]\Omega =2}$ are the $\Omega$-doublet constants. With this, the doublet frequency splitting shown on Fig.~\ref{f:WCDoublet} can be fitted using,

\begin{equation}\label{eq:5}
\delta\nu= \pm 2\tilde{o}_{^3\Delta_1}J(J+1) \pm 2\tilde{o}_{[20.6]\Omega =2}(J-1)J(J+1)(J+2).
\end{equation}
where the $\pm$ signs indicate 4 different possibilities of $\left \{  +,+\right \}$, $\left \{  +,-\right \}$, $\left \{  -,+\right \}$, and $\left \{  -,-\right \}$ for the fit. We note that $\left \{  -,+\right \}$, $\left \{  -,-\right \}$ cases would give the same fit result as the $\left \{  +,+\right \}$, $\left \{  +,-\right \}$ cases, only with the opposite signs for the constants. As we only care about the magnitude of the doublet constants, this reduces the possibilities to 2 cases. The least squared fit for two different cases of $\left \{  +,+\right \}$ and $\left \{  +,-\right \}$ both showed small residuals, and also revealed that $2\tilde{o}_{^3\Delta_1}J(J+1)$ term is much larger than the $2\tilde{o}_{[20.6]\Omega =2}(J-1)J(J+1)(J+2)$ term. In other words, the first term was the main contribution to the observed frequency splitting, that made the second term to be irrelevant whether it was added or subtracted from the first term.

Fitting the $\Omega$-doublet frequency splittings observed in R(4) and R(5) lines of $^{182}$W$^{12}$C, $^{184}$W$^{12}$C, $^{186}$W$^{12}$C isotopes gave the ground state $\Omega$-doublet constant $\tilde{o}_{^3\Delta_1}$, and the upper bound of the excited state $\Omega$-doublet constant $\tilde{o}_{[20.6]\Omega =2}$, which are shown on table~\ref{table:7}. Details of the fitting procedure, and the LIF measurements of $^{182}$W$^{12}$C, $^{186}$W$^{12}$C isotopes are shown in the Appendix. The isotope dependent doublet constants are all within the error bars of our extracted constants. The measured $\Omega$-doublet constant $\tilde{o}_{^3\Delta_1}$ is used for the estimation of the difference in g-factors between the doublet levels, which will be discussed in section IV.

\begin{table}
\caption{The $\Omega$-doublet constants of $^3\Delta_1$ and $[20.6]\Omega=2$ states are shown and compared. Ref.~\cite{Wang2012} shows the $\Omega$-doublet constants of $^3\Delta_1$ for the $^{184}$W$^{12}$C. The 1 $\sigma$ fitting error is shown inside the parentheses for both experimental values.}\label{table:7}

\begin{ruledtabular}
\begin{tabular}{@{}lrr}

$\Omega$-doublet Constant & This Work & Ref.~\cite{Wang2012} \\
               &(kHz)     &(kHz)     \\
\hline
&&\\
$\tilde{o}_{^3\Delta_1}$ &418(18)&400(13)\\
&&\\
$\tilde{o}_{[20.6]\Omega =2}$ & $<1$ & \\
&&\\
\end{tabular}
\end{ruledtabular}
\end{table}

\section{Systematic Uncertainties of WC electron EDM experiment}

\subsection{Systematic Uncertainty from $\mathcal{E_{\textrm{eff}}}$}

In the previous section, we have reported on the hyperfine constants of both ground $^3\Delta_1,v"=0$ state and excited $[20.6]\Omega=2,v'=4$ state. The part that we are interested in is the ground state hyperfine constant $h_{^3\Delta_1,v"=0}$, as it contains the information of the WC electron wavefunction near the tungsten nucleus. As the $\mathcal{E_{\textrm{eff}}}$ field comes from the relativistic effect being applied to the WC valence electrons near the heavy nucleus~\cite{Sandars1966}, the calculation of $\mathcal{E_{\textrm{eff}}}$ in a specific electronic state is closely linked to the calculation of hyperfine constant in the same state. Therefore, the comparison between the calculated WC hyperfine constant $h_{^3\Delta_1,v"=0}$ and the experimental results could serve as a qualitative test of self consistency in the calculation of $\mathcal{E_{\textrm{eff}}}$.

To perform {\it ab~initio} calculation of the WC molecule, a generalized relativistic effective core potential (GRECP)~\cite{Titov1999,Mosyagin2010} for
tungsten atom was generated. As a result, 60 core electrons ($1s{-}4f$) of W were explicitly excluded from the correlation treatment. Basis set for W was
constructed using the generalized correlated scheme~\cite{Mosyagin2001} and it consists of $7 s$-, $9 p$-, $6 d$-, $4 f$- and $2 g-$type contracted
gaussians;
such a basis is usually written as [7,9,6,4,2].  The aug-cc-pVTZ basis set~\cite{Dunning1989} reduced to [5,4,2,1] was used for carbon.

To evaluate $\mathcal{E_{\textrm{eff}}}$, one needs to compute the following parameter of the P,T-odd molecular Hamiltonian (discussed in Refs.~\cite{Kozlov1987, Kozlov1995, Titov2006}):
\begin{equation}
\label{Wd}
W_d = \frac{1}{\Omega d_e}
\langle \Psi|\sum_iH_d(i)|\Psi
\rangle,
\end{equation}
where $\Psi$ is the wave function of the considered state $^3\Delta_1$, and
$\Omega$ is the projection of total
electronic momentum on the molecular axis directed from W to C,
\begin{eqnarray}
H_d=2d_e
\left(\begin{array}{cc}
0 & 0 \\
0 & \bm{\sigma E} \\
\end{array}\right)\ ,
\label{Hpt}
\end{eqnarray}
$\bm{E}$ is the inner molecular electric field, and $\bm{\sigma}$ are the Pauli matrices. In these designations, $\mathcal{E_{\textrm{eff}}}=W_d|\Omega|$.

The hyperfine constant $A_{||}$ for the $^3\Delta_1$ state is
\footnote{for the case of $\Omega=1$$\rightarrow$$A_{||} = h$,
for the case of $\Omega=2$$\rightarrow$$A_{||} = h/4$},
%
\begin{eqnarray}
A_{||}=\frac{\mu_{\rm W}}{I\Omega}
   \langle
   \Psi_{^3\Delta_1}|\sum_i\left(\frac{\bm{\alpha}_i\times
\bm{r}_i}{r_i^3}\right)_z
|\Psi_{^3\Delta_1}
   \rangle
\label{Hhfs}
\end{eqnarray}
where $\mu_{\rm W}=0.11778471 \mu_{\rm N}$~\cite{Mills1993}.

To calculate matrix elements~\ref{Wd} and ~\ref{Hhfs}, we used a two-step scheme, where one performs correlation calculation for the valence and outer
core electrons with the GRECP, followed by the nonvariational one-center restoration of the wave function at the inner core region of W (see~\cite{Titov2006} for more details). Twenty electron correlation calculation was performed using the spin-orbit direct multireference configuration interaction (SODCI) approach~\cite{Alekseyev2004}, accounting for the spin-orbit selection procedure~\cite{Titov2001}. As a basis set of one-electron functions for the SODCI calculation, we used the eigenvectors (natural orbitals) of some one-electron density matrix calculated at the scalar-relativistic coupled-clusters level with single and double cluster amplitudes using the {\sc cfour} code~\cite{CFOUR}.

$\mathcal{E_{\textrm{eff}}}$ and $A_{||}$ were calculated at R(W-C)=3.2248~a.u., which is close to the equilibrium distance. The calculated $A_{||}$ value is $\sim-1192$MHz and the $\mathcal{E_{\textrm{eff}}}$ value is $\sim-36$GV/cm. The influence of interaction with the low-lying electronic state $^3\Delta_2$ on the hyperfine structure will be discuss in the following subsection along with the $\Omega$-doublet analysis.



\begin{table}
\caption{The hyperfine constants of $^{183}W^{12}C$ in $[20.6]\Omega=2, v'=4 \leftarrow X^3\Delta_1,v"=0$ transition is shown and compared. The 1 $\sigma$ fitting error is shown inside the parentheses for both experimental values. The calculated value is believed to be correct within $10\%$ level.}\label{table:5}

\begin{ruledtabular}
\begin{tabular}{@{}lrrr}

Hyperfine Constant & This Work & Ref.~\cite{Wang2011a} & Calculated\\ 
               &(MHz)     &(MHz)     &(MHz)\\
\hline
&&\\
$h_{^3\Delta_1}$     &-1171(4)&-1363(17)&-1192($10\%$)\\
&&\\
$h_{[20.6]\Omega=2}$&1258(6)&&\\
&&\\
\end{tabular}
\end{ruledtabular}
\end{table}

We report the measured hyperfine constants of $h_{^3\Delta_1,v"=0}$ and $h_{[20.6]\Omega=2,v'=4}$ in Table~\ref{table:5}. For the hyperfine constant of ${^3\Delta_1,v"=0}$ state, we compare our result with the previous results of Wang's~\cite{Wang2011a}, which had a slightly larger uncertainty then our case. As seen from Table~\ref{table:5}, our result of $h_{^3\Delta_1}$ had good agreement with the theoretical calculation, however, disagreed with ref.~\cite{Wang2011a} in $10\%$ level. The disagreement was relatively large considering the uncertainties assigned on each experimental hyperfine constants. On the other hand, all three results of $h_{^3\Delta_1}$ constant agrees within $10\%$ level. These results give information on the ground state electronic wave function at the nucleus of the molecule, which is strongly related to the effective electric field calculation of the valence electrons in WC. Therefore, we believe our effective electric field calculations are self consistent at least in $10\%$ level, which is given by the hyperfine constant comparison.

\subsection{Systematic Uncertainty from $\Delta g$}

The biggest systematic uncertainty of the electron EDM measurement with YbF molecules, which holds the current experimental limit, came from the imperfect $\mathcal{E_{\textrm{lab}}}$ reversal combined with the r.f. phase detuning~\cite{Hudson2011}. However, this will not be present in our case. Instead of reversing the $\mathcal{E_{\textrm{lab}}}$ for the Zeeman shift cancelation, our measurement scheme shown on Fig.~\ref{f:WCUncertainty} uses the $\Omega$-doublet levels of $^3\Delta_1$ state~\cite{Lee2009}. Our systematic uncertainty would come from a different source, which is the small difference in g-factors between the top and bottom doublet levels. The upper bound of systematic uncertainty in our electron EDM measurement scheme can be written as $2\mu _{B}B\Delta g$.

As mentioned previously in the introduction section, $\Delta g$ is closely related to the energy splitting between the top and bottom levels of the $\Omega$-doublet~\cite{Petrov2011}. We use the constant $\tilde{o}_{^3\Delta_1}$ reported in table~\ref{table:7} for the calculation of the g-factors of top and bottom levels of the $\Omega$-doublet. We need to consider the influence of interaction with the low-lying electronic state $^3\Delta_2$ on the hyperfine structure and g-factors of the $^3\Delta_1$ state of WC, where rough ten-electron calculations were performed (core states of $5s^25p^6$ of tungsten and $1s^2$ of fluorine were excluded from the correlation treatment). The calculated off-diagonal {\it electronic} matrix elements are,

\begin{eqnarray}
 \label{double2}
	\Delta/2=B'\langle\Psi_{^3\Delta_1}|J^e_-
 |\Psi_{^3\Delta_2}
   \rangle~ = 0.8 ~{\rm cm}^{-1}, \\
 \label{Aperp2}
   \frac{\mu_{\rm W}}{I}
   \langle
   \Psi_{^3\Delta_1}|\sum_i\left(\frac{\bm{\alpha}_i\times
\bm{r}_i}{r_i^3}\right)
_-|\Psi_{^3\Delta_2}
   \rangle~ = 2742 ~{\rm MHz}, \\
 \label{Gperp2}
   G_{\perp} = \langle
   \Psi_{^3\Delta_1}|{J}^e_- +  {S}^e_-
|\Psi_{^3\Delta_2} \rangle = 3.1 .
\end{eqnarray}

The required diagonal {\it electronic} matrix elements and excitation energies are taken from experiment:
$B' = 0.509 ~{\rm cm}^{-1}$, $A_{\parallel} = -1171 ~ {\rm GHz}$ (this work), $G_{\parallel}
= 0.022$ \cite{Wang2011b}, $D = 1.53 ~{\rm a.u.}$ \cite{Wang2011a},
$E_{^3\Delta_2}-E_{^3\Delta_1} = 1194~\rm{cm}^{-1}$ \cite{Rothgeb2008}.

\begin{table}
\caption{
Calculated values of HFS as a
function of $J$ for the $^3\Delta_1$ state of WC.
The second(third) and fifth(sixth) columns are the results obtained without(with)
interaction from the $^3\Delta_2$ state taken into account.}
\begin{center}
\begin{ruledtabular}
\begin{tabular}{cccccccc}

    J   &                &            &    J   &                &            \\
\hline
&\\
    1   &        882.5   &     882.5  &   16   &         72.4   &       42.2 \\
    2   &        495.5   &     492.4  &   17   &         68.2   &       36.2 \\
    3   &        347.5   &     342.2  &   18   &         64.6   &       30.6 \\
    4   &        268.3   &     260.8  &   19   &         61.2   &       25.5 \\
    5   &        218.7   &     209.2  &   20   &         58.2   &       20.6 \\
    6   &        184.6   &     173.2  &   21   &         55.5   &       16.1 \\
    7   &        159.8   &     146.5  &   22   &         53.1   &       11.7 \\
    8   &        140.9   &     125.6  &   23   &         50.8   &        7.6 \\
    9   &        126.0   &     108.8  &   24   &         48.7   &        3.7 \\
   10   &        113.9   &      94.9  &   25   &         46.8   &        0.1 \\
   11   &        104.0   &      83.1  &   26   &         45.1   &       -3.7 \\
   12   &         95.6   &      72.9  &   27   &         43.4   &       -7.2 \\
   13   &         88.5   &      63.9  &   28   &         41.9   &      -10.5 \\
   14   &         82.4   &      55.9  &   29   &         40.5   &      -13.8 \\
   15   &         77.1   &      48.7  &   30   &         39.1   &      -17.0 \\
&\\
\end{tabular}
\label{hfs}
\end{ruledtabular}
\end{center}
\end{table}

In table~\ref{hfs}, the hyperfine splittings (HFS) calculated from E($F=J-1/2$)$-$E($F=J+1/2$) are given for each rotational levels of the $^3\Delta_1$
state of WC.  The hyperfine structure results were obtained by numerical diagonalization of the Hamiltonian, which was described in ref.\cite{Petrov2011}.  Interaction with the $^3\Delta_2$ state equally influences the properties of the $|e\rangle$ and $|f\rangle$ states, whereas the interactions with $0^+$ and $0^-$ states would lead to the different properties between the $|e\rangle$ and $|f\rangle$ states. However, the influence of the latter interactions on HFS is not considered in the present study due to the complexity of the WC spectrum. The experimental value of the $\Omega$-doubling allows us to write,

\begin{eqnarray}
&\nonumber
 B'^2 \left( \sum_n\frac{\langle\Psi_{^3\Delta_1}|J^e_+
 |\Psi_{n0^+}\rangle^2}{E_{^3\Delta_1} - E_{n0^+}} -
  \sum_m\frac{\langle\Psi_{^3\Delta_1}|J^e_+
\label{doubl}
 |\Psi_{m0^-}\rangle^2}{E_{^3\Delta_1} - E_{m0^-}} \right) \\
&= \pm 0.4~~\rm{MHz}.
\end{eqnarray}

Due to a large number of electronic states in the WC spectrum, it is hard to reproduce this value in {\it ab~initio} calculations. However, we have estimated
from our calculations that the matrix elements
$\langle\Psi_{^3\Delta_1}|S^e_+ |\Psi_{n0^\pm}\rangle $
 is much smaller than
$\langle\Psi_{^3\Delta_1}|J^e_+ |\Psi_{n0^\pm}\rangle $ by absolute value.
Therefore, we can write,

\begin{eqnarray}
\label{doublg}
  \sum_n\frac{2B'\langle\Psi_{^3\Delta_1}|J^e_+|\Psi_{n0^+}\rangle
 \langle\Psi_{^3\Delta_1}|J^e_+ + S^e_+|\Psi_{n0^+}\rangle}
 {E_{^3\Delta_1} - E_{n0^+}} -
\\
\nonumber
 \sum_m\frac{2B'\langle\Psi_{^3\Delta_1}|J^e_+|\Psi_{m0^-}\rangle
  \langle\Psi_{^3\Delta_1}|J^e_+ + S^e_+ |\Psi_{m0^-}\rangle}
  {E_{^3\Delta_1} - E_{m0^-}}  \\
\nonumber
  \approx \pm 5 \cdot 10^{-5}.
\end{eqnarray}

The $\pm$ sign in eqs.~(\ref{doubl}) and (\ref{doublg}) comes from the fact that it has not been experimentally determined whether it is $|e\rangle$ or $|f\rangle$ state that belongs to the lower $\Omega$-doublet level. Taking this into account, the difference of the g-factors can be written as $|g_e-g_f| = 5 \cdot 10^{-5} \cdot J(J+1)$. It can also be shown that interaction with the $^3\Delta_2$ state leads to the same $J-$dependence of the g-factors for $|e\rangle$ and $|f\rangle$ levels:
$g_{e(f)}=0.022+\Delta G_{\perp}/2
(E_{^3\Delta_2}-E_{^3\Delta_1})
   \cdot(J+2)(J-1)$.

\begin{figure}
\includegraphics[width = 3.3 in]{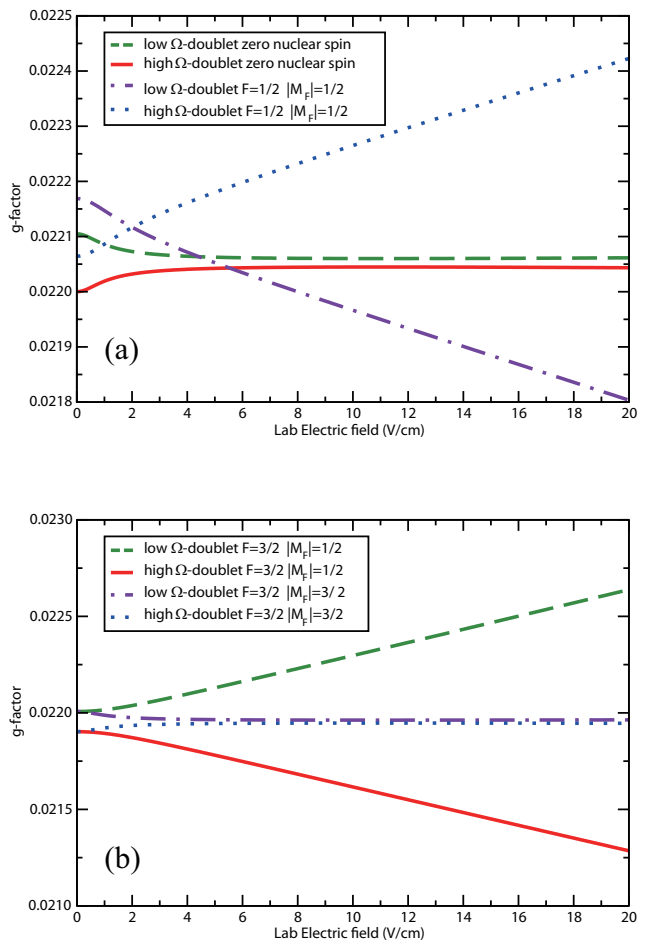}
\caption{(a)Calculated g-factor curves for $|J=1, F=1/2, |M_F|=1/2\rangle$ hyperfine levels of $^{183}$W$^{12}$C, and $J=1$ rotational levels of WC isotope with zero nuclear spin. (b)Calculated g-factor curves for $|J=1, F=3/2, |M_F|=1/2\rangle$ hyperfine levels of $^{183}$W$^{12}$C, and $|J=1, F=3/2, |M_F|=3/2\rangle$ hyperfine levels of $^{183}$W$^{12}$C}
\label{gfgecross}
\end{figure}

Smaller the difference between $g_e$ and $g_f$ is, smaller the systematics would be coming from spurious magnetic fields. This difference depends on the electric field, which are shown on Fig.~\ref{gfgecross}. Let us analyze the behavior of g-factors in this plot. At first glance, one expects that the difference between $g_e$ and $g_f$ could be made zero by increasing the electric field, since the external electric field mixes $|e\rangle$ and $|f\rangle$ levels.
However, this only happens for $|J=1, F=1/2, |M_F|=1/2\rangle$ hyperfine levels of $^{183}$W$^{12}$C, which will be explained in subsection C. Without a proper perturbation term from the hyperfine interaction, the difference between g-factors would eventually diverge as the electric field increases, which has been observed from PbO molecules~\cite{Bickman2009}. This behavior was explained by M.G.\ Kozlov(see acknowledgements in \cite{Bickman2009}).

Applying the same analysis to the case of WC molecules, when $J=2$ level is mixed in to $J=1$ by the external electric field, the g-factor of the lower $\Omega$-doublet level will have a slight decrease at very low electric field, and continuously increase as the field increases (here we suggest that electronic g-factor is positive). At the absence of external electric field, it was found that the initial value for g-factor of the lower $\Omega$-doublet level is larger than the g-factor of the higher one~\cite{Kawall2004}. Taking these two factors into account, one could understand the g-factor curves in Fig.~\ref{gfgecross}(a) for isotopes with zero nuclear spin. Without the hyperfine interaction, the difference between g-factors would eventually diverge at higher electric fields, however, was not shown in Fig.~\ref{gfgecross}(a) due to limited plot range for comparison purposes.

It can also be shown that whenever signs of the sums in Eqs. (\ref{doubl}) and (\ref{doublg}) are the same, the lower $\Omega$-doublet level will have a larger g-factor (independently from the parity of its state).
This is true for WC since
$\langle\Psi_{^3\Delta_1}|S^e_+ |\Psi_{n0^\pm}\rangle $
 is much smaller than
$\langle\Psi_{^3\Delta_1}|J^e_+ |\Psi_{n0^\pm}\rangle $ by absolute value.
Therefore, we arrive at the same conclusion of g-factors diverging at high electric field, for WC isotopes with zero nuclear spin.

%
On Fig.~\ref{gfgecross} the calculated g-factors for J=1 $\Omega$-doublet levels
of $^3\Delta_1$ state of WC are given as functions of electric field.
Here, for simplicity we assume that the value of $g = 0.022$ \cite{Wang2011b}
belongs to a high $\Omega$-doublet
level of spinless isotopes.
Further experimental investigation would be required to confirm this.
However, even if it were the opposite case, it would only shift all the curves in the figure slightly down, leaving their relative positions unchanged.

\begin{figure}
\includegraphics[width = 3.3 in]{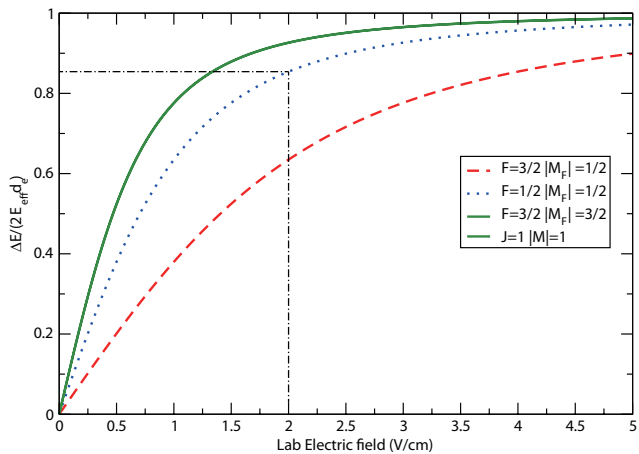}
\caption{Calculated EDM induced Stark splitting between $\pm M_F$ levels
of the $|J=1, F=1/2\rangle$; $|J=1, F=3/2\rangle$ and $J=1$ states. All the curves are normalized by the maximum EDM induced Stark
splitting, $2 \mathcal{E_{\textrm{eff}}}\cdot d_e$. The dot-dash line shows the EDM induced Stark splitting between $\pm M_F$ levels of $|J=1, F=1/2\rangle$ state at $\mathcal{E_{\textrm{lab}}}=2V/cm$.}
\label{ssplit}
\end{figure}

The maximum EDM induced Stark splitting $2 \mathcal{E_{\textrm{eff}}}\cdot d_e$ is only reached for fully polarized molecule. For the finite electric field, the Stark splitting obtained from numerical calculations is plotted in Fig.~\ref{ssplit}.

%

As shown on Fig.~\ref{gfgecross}(a), we can get the $\Delta g = 0.00002$ for isotope with zero nuclear spin at $\mathcal{E_{\textrm{lab}}}=10$V/cm, which is enough laboratory electric field to fully polarize WC molecule. Using this estimation of $\Delta g$, we get $\sim100\mu$Hz of systematic shift with $\sim\mu$G control of the magnetic field. This would limit our EDM sensitivity at $\delta d_e \approx 10^{-29}$ e-cm level when $\mathcal{E_{\textrm{eff}}}\sim-36$GV/cm.

\subsection{Further Suppression of Systematic Uncertainty from $\Delta g$}

There is an additional interesting feature shown on Fig.~\ref{gfgecross}(a), where the g-factors cross at $\mathcal{E_{\textrm{lab}}}=2V/cm$ for $|J=1, F=1/2, |M_F|=1/2\rangle$ hyperfine levels of $^{183}$W$^{12}$C isotope due to hyperfine interaction. For $|J=1, F=1/2, |M_F|=1/2\rangle$ and $|J=1, F=3/2, |M_F|=1/2\rangle$ states, the perturbing state is the nearest hyperfine energy level, which is simply each other ($|J=1, F=3/2, |M_F|=1/2\rangle$ is the perturbing state for $|J=1, F=1/2, |M_F|=1/2\rangle$ and vice versa). Note that for $A_{||} < 0 $, $|J=1, F=1/2, |M_F|=1/2\rangle$ level is higher than $|J=1, F=3/2, |M_F|=1/2\rangle$ level. Therefore, the energy denominator for $|J=1, F=1/2, |M_F|=1/2\rangle$ levels in the perturbation theory will have the opposite sign compared to the molecule without hyperfine structure. Accordingly, if the mixing by the external electric field of the $|J=1, F=3/2, |M_F|=1/2\rangle$ level is taken into account, the lower $\Omega$-doublet level of the $|J=1, F=1/2, |M_F|=1/2\rangle$ will have smaller g-factor with increasing electric field. As the initial value for g-factor of the lower $\Omega$-doublet level is larger than the g-factor of the higher $\Omega$-doublet level, the corresponding curves for $g_e$ and $g_f$ are crossed. On the other hand, for $|J=1, F=3/2, |M_F|=1/2\rangle$ level, the perturbing state is the higher lying $|J=1, F=1/2, |M_F|=1/2\rangle$ state. Therefore the g-factor curves shown in Fig.~\ref{gfgecross}(b) does not crossed for the $|J=1, F=3/2, |M_F|=1/2\rangle$ level. The perturbing state of $|J=1, F=3/2, |M_F|=3/2\rangle$ level is the hyperfine state of the much higher lying $J=2$ level, which is the $|J=2, F=3/2, |M_F|=3/2\rangle$ state. As a result, the g-factor curves of $|J=1, F=3/2, |M_F|=3/2\rangle$ level in Fig.~\ref{gfgecross}(b) resembles the g-factor curves for isotopes with zero nuclear spin shown in Fig.~\ref{gfgecross}(a).

Utilizing the g-factor crossing point of the $|J=1, F=1/2, |M_F|=1/2\rangle$ hyperfine levels of $^{183}$W$^{12}$C isotope, one could further suppress systematic uncertainty coming from $\Delta g$. There are two main factors to check before we could consider using this crossing point for the electron EDM measurement.

The most important factor to check is, how much of $\mathcal{E_{\textrm{eff}}}$ field we would get at the the g-factor crossing point with the given $\mathcal{E_{\textrm{lab}}}$. As the crossing point lies at $\mathcal{E_{\textrm{lab}}}=2$V/cm, which is not enough laboratory electric field to fully polarize the WC molecule, we would not get the full $\mathcal{E_{\textrm{eff}}}\sim-36$GV/cm. However, as shown on Fig.~\ref{ssplit}, $\mathcal{E_{\textrm{lab}}}=2$V/cm applied to the $|J=1, F=1/2, |M_F|=1/2\rangle$ state would still give $85\%$ of EDM induced Stark splitting, compared to the fully polarized case. In other words, we only loose $15\%$ of the $\mathcal{E_{\textrm{eff}}}$ field, and in return we get effectively zero difference in g-factors, which would suppress the systematics even further.

The second factor to consider is the slope of the g-factor curves at the crossing point. As we have limited $\mathcal{E_{\textrm{lab}}}$ stability, the uncertainty in $\mathcal{E_{\textrm{lab}}}$ would get transferred to the uncertainty in g-factors at the crossing point. By simple calculation of, 2$\times$  ``$\mathcal{E_{\textrm{lab}}}$ field stability"$\times$``the slope of the g-factor curve at the crossing point", we get $\Delta g$ of $0(5)\times10^{-8}$, where the number inside the parenthesis shows the uncertainty, with $\sim$mV/cm control of the $\mathcal{E_{\textrm{lab}}}$ field. Combining the above two factors, we expect the systematic uncertainty of electron EDM measurement due to $\Delta g$ to be 1000 times smaller than case explained in previous subsection B.

There is a minor drawback of this EDM measurement scheme in terms of statistics. One is due to relatively low natural abundance of $14\%$ for the $^{183}$W isotope, which is about half of the most abundant $^{184}$W isotope with zero nuclear spin. The other drawback comes from a relatively small Clebsch-Gordan coefficient for the hyperfine transition. As discussed previously, the intensity ratios of the hyperfine transitions in R(1) line follows $[I_a : I_b : I_c = 9 : 1 : 5]$, and the $|J=1, F=1/2\rangle$ EDM state can only couple to $|J=2, F=3/2\rangle$ excited state, which corresponds to the c transition, the second strongest. With these two factors, the electron EDM measurement scheme using the g-factor crossing points of $|J=1, F=1/2, |M_F|=1/2\rangle$ hyperfine levels of $^{183}$W isotope would suffer factor of 2 loss in statistics compared with the EDM measurement with spinless isotope. However, this is a small loss compared to the factor of 1000 gain in systematic uncertainty, therefore would become a useful alternative electron EDM measurement scheme with enhanced comagnetometer performance.

\section{Statistical Sensitivity of WC electron EDM experiment}

Due to the reasons explained in the previous section, we believe the WC electron EDM measurement would be limited by the statistics rather than the systematic uncertainties at this point. In this section, we analyze the beam properties in order to show the achievable level of statistical sensitivity limit of the electron EDM measurement, and also discuss about possible ways to improve this limit.

\subsection{Beam Properties Related to Statistics}

The spectrum of WC molecule gives multiple information about the beam properties. The height of each peaks in Fig.~\ref{f:183WCHyper} and ~\ref{f:WCDoublet} represent the molecular flux at a given electronic, vibrational, and rotational quantum state. Conversely, knowing the quantum states and the scattering rate of the transition, we can estimate the density of molecules within the detection volume. The Doppler broadened linewidth can be used for calculation of the beam divergence, and the axial velocity can be calculated from the time delay between the triggering of the gas valve and the PMT gate.

\begin{figure}
\includegraphics[width = 3.3 in]{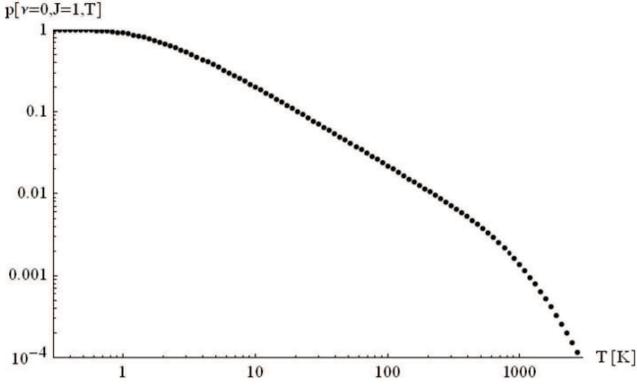}
\caption{Fractional ro-vibrational ground state $(v=0, J=1)$ population at different temperatures ranging from $0.3$K to $3000$K. The fraction converges to 1 as the temperature goes below 1K.}
\label{f:rovibe}
\end{figure}

As the proposed electron EDM measurement scheme uses the WC molecules in their rotational and vibrational ground state of the $X^3\Delta_1$ state~\cite{Lee2009}, estimation of fractional rotational ground state population becomes important. Intuitively, more molecules would be in their rovibrational ground state as the internal temperature of the beam cools down by the buffer gas collision. From statistical physics, the probability of rotational ground state population within the X$^3\Delta_1,v"=0$ state at temperature T can be calculated from the following equation,

\begin{equation}\label{eq:4-1}
P(J",T) =\frac{(2J"+1)Exp[-\frac{h}{kT}[BJ"(J"+1)]]}{\sum_{J=0}^{J_{max}} (2J+1)Exp[-\frac{h}{kT}[BJ(J+1)]]},
\end{equation}
where $B$ is the ground state rotational constant given by ref.~\cite{Sickafoose2002}. Figure~\ref{f:rovibe} shows the fractional rovibrational ground state $(v'=0, J'=1)$ population at different temperatures ranging from $0.3$K to $3000$K. From Fig.~\ref{f:183WCHyper} and~\ref{f:WCDoublet}, we observed a general trend of R(1) lines being stronger than the R(2) lines. This condition combined with equation~\ref{eq:4-1} gives an upper rotational temperature limit of $T_{rot}<5.5k$. Therefore, we expect minimum of $35\%$ of WC molecules to be at their rovibrational ground state when $T_{rot}=5.5k$. For better estimation of the rotational temperature, accurate ratios among multiple rotational lines are required.

All the parameters mentioned above are listed on table~\ref{table:3}. We separate the directly measured quantities from the calculated quantities. In particular, the Frank Condon factor has been estimated based on dispersed fluorescence spectroscopy of the same transition~\cite{Sickafoose2002}. The number of WC molecules within the detection volume has been estimated based on the geometric efficiency, quantum efficiency, scattering rate of the transition, and the Frank Condon factor. These parameters will be used for calculation of the statistical sensitivity limit of electron EDM experiment in the following subsection.

\begin{table}
\caption{The list of measured and calculated parameters related to the WC molecular beam.}\label{table:3}

\begin{ruledtabular}
\begin{tabular}{@{}lrr}

Parameter & Measured & Calculated \\
\hline
&&\\

Beam Divergence&$0.00041 sr$&\\
&&\\
Beam Axial Velocity&$650 m/s$&\\
&&\\
Distance Between&$75 cm$&\\
Source and Probe&&\\
&&\\
Geometric Photon&&$0.003$\\
Collection Efficiency&&\\
&&\\
Quantum Efficiency of&&$0.1$\\
Photo Multiplying Tube&&\\
&&\\
Franck Condon Factor&&$0.0001$\\
of $v'=4 \leftarrow v"=0$ excitation&&\\
and $v'=4 \rightarrow v"=0$ decay&&\\
&&\\
WC Molecules&&$\sim10^{7}$WC/pulse\\
within detection volume&&\\
&&\\
$T_{rot}$&$<5.5K$&\\
&&\\
Fractional Ro-vibrational&&$>35\%$\\
Ground State &&\\
&&\\

\end{tabular}
\end{ruledtabular}
\end{table}

\subsection{Current Level of Statistical Uncertainty and Possible Improvements}

The EDM of an electron, $d_e$, inside a fully polarized WC molecule would produce a frequency shift of $\Delta \nu_{\textrm{EDM}} = 2d_e\mathcal{E_{\textrm{eff}}}/h$. The statistical uncertainty of the EDM shift is defined by the Fourier-limited frequency resolution of $\Delta \nu_{\textrm{stat}} = 1/{2\pi \tau \sqrt{\dot{N}\times T}}$, where $\tau$ is the interrogation time and $\dot{N}$ is the rate of measurement, and $T$ is the duration of measurement. Therefore, we can write down the expression for statistical uncertainty in the measurement of $d_e$ as,

\begin{equation}\label{eq:4-2}
\delta d_e = \frac {h}{4\pi \tau \mathcal{E_{\textrm{eff}}} \sqrt{\dot{N}\times T}}.
\end{equation}

From eq.~\ref{eq:4-2}, longer interrogation time, larger effective electric field, and more number of measurements result in smaller statistical uncertainty of $d_e$. Going back to Table~\ref{table:3}, one can make the connections between the beam properties and the parameters in eq.~\ref{eq:4-2}. Slower beam axial velocity would give longer interrogation time at fixed distance. Lower beam divergence would increase the molecular density within the LIF probe laser volume, which would result in higher photon scattering rate. Colder rotational temperature would also enhance the photon scattering rate as more molecules would be in their rovibrational ground state.

The list of measured and calculated parameters in eq.~\ref{eq:4-2} are shown on Table~\ref{table:8}. The current status of the experiment gives a projected statistical uncertainty of $\delta d_e = 5\times10^{-27} e-cm/\sqrt{Day}$. From our systematic uncertainty analysis, we believe this would be the dominant term that limits our electron EDM measurement sensitivity, which is slightly above the current experimental limit~\cite{Hudson2011}.

\begin{table}
\caption{The list of measured and calculated parameters related to electron EDM experiment.}\label{table:8}

\begin{ruledtabular}
\begin{tabular}{@{}lrr}

Parameter & Measured & Calculated \\
\hline
&&\\

Interrogation Time ($\tau$)&$1.25 ms$&\\
&&\\
Photon Count Rate ($\dot{N}$)&$5 photons/s$&\\
&&\\
Effective Electric Field ($\mathcal{E_{\textrm{eff}}}$)&&$-36GV/cm$\\
&&\\
Projected $\delta d_{e,stat}$ && $5\times10^{-27} \frac {e-cm}{\sqrt{Day}}$\\
&&\\
\end{tabular}
\end{ruledtabular}
\end{table}

One possible improvement on this limit could come from extension of the WC beam line to increase the interrogation time. Due to the EDM sensitive $X^3\Delta_1$ state being in the ground state of WC, our interrogation time is only limited by the time of flight of the molecules, not by the lifetime of the state. As seen on Table~\ref{table:3}, our beam has very low divergence, therefore extending the beam line of the molecules could be an option for us in order to increase the time of flight. From a conservative estimate, we believe factor of $2\sim3$ of beam line extension is plausible without the lost of fluorescence rate.

Another prospect of improvement for statistics comes from the Franck Condon factor. From the dispersed fluorescence spectroscopy of ref.~\cite{Morse2011}, it was shown that $v'=4 \leftarrow v"=0$ vibrational excitation of $[20.6]\Omega=2 \leftarrow X^3\Delta_1$ transition happens at $\sim100$ times lower rate than $v'=1 \leftarrow v"=0$ vibrational excitation of the same electronic transition. Therefore, probing a new stronger transition of $[20.6]\Omega=2, v'=1 \leftarrow X^3\Delta_1,v"=0$ at a known wavelength of 545nm~\cite{Sickafoose2002}, could give us a factor of 100 gain in $\dot{N}$. Combining the gains from the vacuum system and the laser system, the statistical uncertainty of $d_e$ is expected to reach $10^{-28} e-cm/\sqrt{Day}$ level in the near future.

\section{Conclusion}

In this work, we have investigated the electron EDM measurement scheme using the $X^3\Delta_1$ ground state of WC molecules, where the focus was on identifying and estimating the main systematic and statistical uncertainties. Well collimated and internally cold beam of WC was developed for electron EDM search in X$^3\Delta_1$ of the molecule. We have shown high resolution LIF spectroscopy of $[20.6]\Omega=2, v'=4 \leftarrow X^3\Delta_1,v"=0$ transition to extract molecular constants relevant to the electron EDM measurement scheme.

The ground state hyperfine constant, $h_{3\Delta_1,v"=0}$, was measured and compared with theoretical calculations to show self consistency in effective electric field calculations at $10\%$ level. The ground state $\Omega$-doublet constant, $\tilde{o}_{^3\Delta_1}$, was measured to give the g-factor curves for top and bottom doublet levels, and also allowed us to estimate the systematic uncertainty coming from imperfect Zeeman shift cancelation in our EDM measurement scheme. The result leads to systematics of $\delta d_e \approx 10^{-29} e-cm$ level with $\sim\mu$G control of the magnetic field. Also an alternative EDM measurement scheme using the g-factor crossing point of hyperfine levels of $^{183}$W$^{12}$C isotope was shown, where the systematics could be suppressed even further. This crossing point of g-factors only happens with molecular isotope with non-zero nuclear spin.

As for the statistical uncertainty, several beam properties such as the flux, velocity, divergence, and rotational temperature were characterized. These properties were linked to the Fourier-limited frequency resolution of the EDM induced Stark shift measurement. The current statistical sensitivity limit was calculated to be $\delta d_e = 5\times10^{-27} e-cm/\sqrt{Day}$. Possible improvements coming from extension of the beam line and the Frank Condon factor were discussed.

We can summarize the main attractive features of WC electron EDM experiment as, (i) verified calculation of large $\mathcal{E_{\textrm{eff}}}$, (ii) use of internal comagnetometer with $\Omega$-doublet structure, (iii) further suppression of systematics due to g-factor crossing point, and (iv) unlimited lifetime of the EDM sensitive state, $X^3\Delta_1$. The former two features are shared by two other molecule based EDM experiments, which are the $^3\Delta_1$ metastable states of thorium oxide(ThO)~\cite{Vutha2010} and hafnium fluoride ion(HfF$^+$)~\cite{Leanhardt2011}. However, the later two features are unique properties of the WC system. Based on our analysis, we believe WC system has many advantages for electron EDM measurement including some unique properties showing a lot of promise for future generation experiments.

\begin{acknowledgments}

J.L. thank M. Morse for giving us valuable comments.
Funding for the work at PNPI was provided by Russian Ministry of Education and
Science, contract \#07.514.11.4141 (2012-2013). L.S.\ is also grateful to the
Dmitry Zimin ``Dynasty'' Foundation.
The molecular calculations were performed at the Supercomputer ``Lomonosov''.

\end{acknowledgments}

\appendix*

\section{$\Omega$-doublet fitting}

\begin{figure}
\begin{center}
\includegraphics[width = 3.3 in]{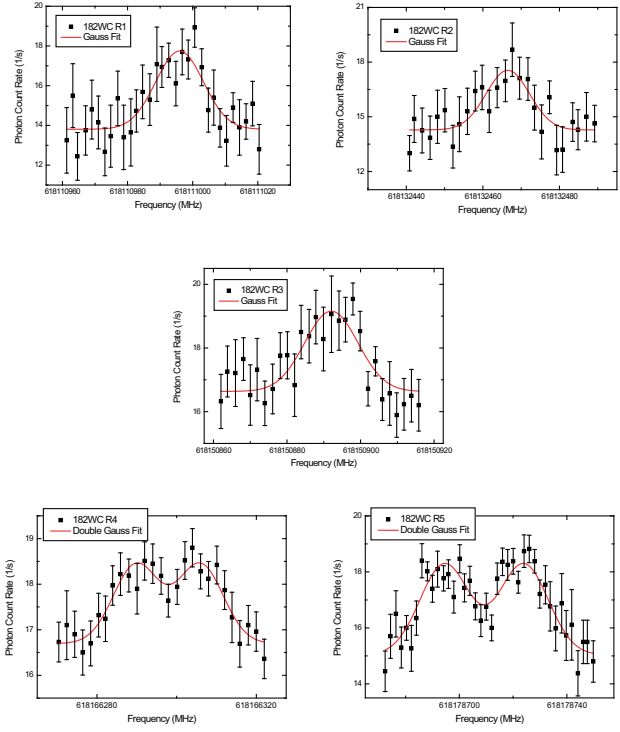}
\caption{R(1) $\sim$ R(5) lines of $[20.6]\Omega=2, v'=4 \leftarrow X^3\Delta_1,v"=0$ transition for $^{182}$W$^{12}$C are shown.}
\label{f:182WCRlines}
\end{center}
\end{figure}

\begin{figure}
\begin{center}
\includegraphics[width = 3.3 in]{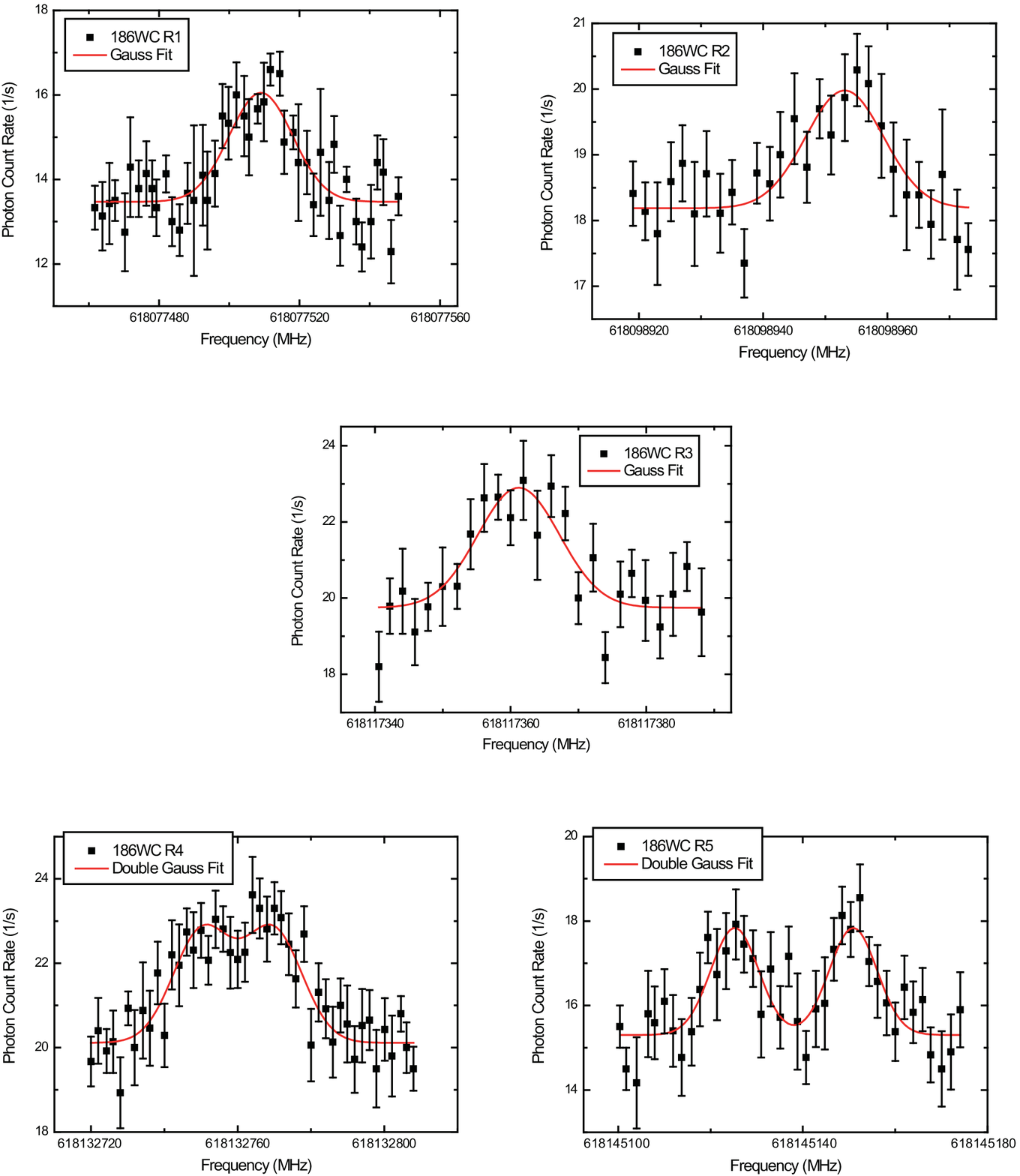}
\caption{R(1) $\sim$ R(5) lines of $[20.6]\Omega=2, v'=4 \leftarrow X^3\Delta_1,v"=0$ transition for $^{186}$W$^{12}$C are shown.}
\label{f:186WCRlines}
\end{center}
\end{figure}

\begin{table}
\caption{Measured $\Omega$-doublet splittings of $^{182}$W$^{12}$C, $^{184}$W$^{12}$C, $^{186}$W$^{12}$C isotopes in R(4) and R(5) lines of $[20.6]\Omega=2, v'=4 \leftarrow X^3\Delta_1,v"=0$ transition. The 1 $\sigma$ errors are shown inside the parentheses in the order of first parenthesis with the statistical uncertainty, and the second parenthesis with the systematic uncertainty in relative frequencies. The residuals from the fits are shown on the third column for the $\left \{  +,+\right \}$ case of eq.~\ref{eq:5}, and the fourth column for the $\left \{  +,-\right \}$ case.}\label{table:6}

\begin{ruledtabular}
\begin{tabular}{@{}lrrr}

Measured Line &Splitting & Fit Residual                 & Fit Residual \\
              &          & for $\left \{  +,+\right \}$ & for $\left \{  +,-\right \}$  \\
               &(MHz)     &(MHz)                  &(MHz)\\
\hline
&&\\
$^{182}$W$^{12}$C, R(4) &16(1.6)(0.01)&-1.4&-1.7\\
$^{182}$W$^{12}$C, R(5) &28(1.6)(0.02)&1.3 &1.5 \\
$^{184}$W$^{12}$C, R(4) &18(1.6)(0.01)&0.7 &0.4\\
$^{184}$W$^{12}$C, R(5) &27(1.6)(0.02)&0.3 &0.5\\
$^{186}$W$^{12}$C, R(4) &18(1.6)(0.01)&0.7 &0.4\\
$^{186}$W$^{12}$C, R(5) &25(1.6)(0.02)&-1.6&-1.4\\
&&\\
\end{tabular}
\end{ruledtabular}
\end{table}

The R branches of $[20.6]\Omega=2, v'=4 \leftarrow X^3\Delta_1,v"=0$ transition of $^{182}$W$^{12}$C and $^{186}$W$^{12}$C isotopes are shown on Fig.~\ref{f:182WCRlines} and ~\ref{f:186WCRlines}. The $\Omega$-doublet splittings observed in R(4) and R(5) lines of these two isotopes combined with the results of $^{184}$W$^{12}$C isotope shown in the Fig.~\ref{f:WCDoublet} are listed on the second column of table~\ref{table:6}.

We couldn't extract any isotope dependence of the $\Omega$-doublet constant from our data as we did not have enough precision on relative frequency shift measurement of the doublet splittings. The isotope dependence of the $\Omega$-doublet constant comes from the difference in reduced mass, which has a fractional difference in $\sim0.1\%$ level~\cite{Wang2012}, however, our measured frequency splittings have fractional uncertainties in $\sim10\%$ level, which were dominated by the fitting errors. This is why we had to fit the measured $\Omega$-doublet splittings from all three isotopes using the same fit function of eq.~\ref{eq:5} and report the isotope independent ground state $\Omega$-doublet constant $\tilde{o}_{^3\Delta_1}$ with a relatively large error bar, and only put an upper bound of 1kHz for the isotope independent excited state $\Omega$-doublet constant $\tilde{o}_{[20.6]\Omega =2}$. Nevertheless, the extracted $\tilde{o}_{^3\Delta_1}$ is consistent with the previous measurement, which was shown on table~\ref{table:7}.

\bibliography{JLeeTungstenCarbide}

\end{document}